\documentclass[pra,twocolumn,superscriptaddress,aps,amsmath,amssymb,nofootinbib]{revtex4}
\usepackage{graphicx}
\usepackage{dcolumn}
\usepackage{bm}
\usepackage{color}

\usepackage[section]{placeins}
\usepackage{graphicx,epsfig}
\usepackage{amsmath}
\usepackage{amsfonts}
\usepackage{braket}
\usepackage{fancyhdr}
\usepackage{hyperref}
\usepackage[utf8]{inputenc}
\usepackage[T1]{fontenc}
\usepackage{amsthm}
\usepackage{hhline}
\usepackage{multirow}
\usepackage{xcolor}
\def\beq{\begin{equation}}
\def\eeq{\end{equation}}
\def\beqa{\begin{eqnarray}}
\def\eeqa{\end{eqnarray}}

\begin{document}

\title{Few-boson localization in a continuum with speckle disorder}

\date{\today}
\author{Pere Mujal}
\affiliation{Departament de F\'{i}sica Qu\`{a}ntica i Astrof\'{i}sica,
Universitat de Barcelona, Mart\'{i} i Franqu\`{e}s 1, 08028 Barcelona, Spain}
\affiliation{Institut de Ci\`{e}ncies del Cosmos (ICCUB), Universitat 
de Barcelona, Mart\'{i} i Franqu\`{e}s 1, 08028 Barcelona, Spain}
\author{Artur Polls}
\affiliation{Departament de F\'{i}sica Qu\`{a}ntica i Astrof\'{i}sica,
Universitat de Barcelona, Mart\'{i} i Franqu\`{e}s 1, 08028 Barcelona, Spain}
\affiliation{Institut de Ci\`{e}ncies del Cosmos (ICCUB), Universitat de 
Barcelona, Mart\'{i} i Franqu\`{e}s 1, 08028 Barcelona, Spain}
\author{Sebastiano Pilati}
\affiliation{School of Science and Technology, Physics Division, 
Universit{\`a}  di Camerino, 62032 Camerino (MC), Italy}
\author{Bruno Juli\'{a}-D\'{i}az}
\affiliation{Departament de F\'{i}sica Qu\`{a}ntica i Astrof\'{i}sica,
Universitat de Barcelona, Mart\'{i} i Franqu\`{e}s 1, 08028 Barcelona, Spain}
\affiliation{Institut de Ci\`{e}ncies del Cosmos (ICCUB), Universitat 
de Barcelona, Mart\'{i} i Franqu\`{e}s 1, 08028 Barcelona, Spain}
\affiliation{Institut de Ci\`{e}ncies Fot\`{o}niques, Parc Mediterrani 
de la Tecnologia, 08860 Barcelona, Spain}
\begin{abstract}
The disorder-induced localization of few bosons interacting via a contact 
potential is investigated through the analysis of the level-spacing statistics 
familiar from random matrix theory. The model we consider is defined in a 
continuum and describes one-dimensional bosonic atoms exposed to the 
spatially correlated disorder due to an optical speckle field. 
First, we identify the speckle-field intensity required to observe, in the single-particle 
case, the Poisson level-spacing statistics, which is characteristic of localized quantum systems, in a computationally 
and experimentally feasible system size. Then, we analyze the two-body 
and the three-body systems, exploring a broad interaction range, from the 
noninteracting limit up to moderately strong interactions. 
Our main result is that the contact potential does 
not induce a shift towards the Wigner-Dyson level-spacing statistics, which would indicate the emergence of an ergodic chaotic state, 
indicating that localization can occur also in interacting few-body systems in a continuum. We also analyze how 
the ground-state energy evolves as a function of the interaction strength.
\end{abstract}
\maketitle
\section{Introduction}
\label{intro}

Since Anderson's 1958 seminal article~\cite{Anderson}, it is known that quenched 
disorder can induce localization of noninteracting quantum particles, determining 
the absence of transport of any conserved quantity in macroscopic samples.
If and when Anderson localization can be stable against inter-particle interactions 
has been an outstanding open question ever since~\cite{Fleishman, Fisher, Altshuler1997}.
In recent years, this question has been addressed in innumerable theoretical 
articles, putting forward the theory of so-called many-body 
localization~\cite{Gornyi,Basko}. This phenomenon is expected to occur in 
isolated one-dimensional systems with disorder. Among 
other properties, it is characterized by the occurrence of perfect insulating 
behavior at finite temperature, by the fact that the many-body localized 
system is unable to act as its own thermal bath, therefore violating 
the eigenstate thermalization hypothesis, and by the emergence of an extensive number 
of local
 integrals of motion. See Refs.~\cite{NandkishoreReview, AletReview} 
for recent reviews.
While some previous theoretical predictions on many-body localization, 
based mostly on perturbative calculations, considered continuous-space 
models~\cite{Gornyi,Basko,Aleiner,Bertoli}, most numerically-exact simulations 
considered one-dimensional discrete-lattice models within the tight binding 
formalism. 
In fact, whether many-body localization can occur in a continuum is still 
a controversial issue.
In Ref.~\cite{Nandkishore}, it is claimed that many-body localization can 
occur even in continuous-space systems if the (non-deterministic) disorder 
is in the impurity limit, but it might be unstable if the correlation 
length of the disorder is finite. Ref.~\cite{Gornyi2017}, instead, states 
that many-body localization cannot occur at all in a continuum.
On the other hand, the continuous-space simulations of Ref.~\cite{Ancilotto}, 
which considered fermionic atoms in a quasi-periodic (hence, deterministic) 
potential and were based on time-dependent density functional theory within 
the adiabatic approximation, displayed one of the experimental hallmarks of 
many-body localization, namely the long-time persistence of an initially 
imprinted density pattern. This phenomenon has indeed been observed in the 
cold-atom experiments on many-body localization~\cite{Schreiber,Luschen,Choi}.
Also a very recent theoretical study states that many-body localization can be 
identified in continuous-space (two-dimensional) systems if the appropriate 
criterion is adopted~\cite{Bertoli2}.
Resolving this controversy is essential, given that discrete-lattice models 
are at most a reasonably good low-energy approximation of experimental systems. 
In this Article, we shed some light on this issue, considering however 
a few-body setup.

The model we consider is tailored to describe a setup that can be implemented 
in cold-atom experiments~\cite{Roati,Billy}. Specifically, it describes bosonic 
atoms in a one-dimensional continuum, interacting via a repulsive zero-range 
interaction. The atoms are exposed to the spatially correlated random potential 
corresponding to the disorder pattern that is generated when an optical speckle 
field is shone onto the atomic cloud.
Due to the higher computational cost of continuous-space models compared to 
lattice models -- for which one could simulate around $10$ itinerant particles or, 
say, $20$ or $30$ immobile spins -- we focus on one-, two-, and three-boson systems. 
The main goal of our analysis is to verify whether localization 
is stable against the repulsive contact inter-particle interaction, meaning 
that many-body localization can be observed in a few-body setup. Indeed, 
many-body localization has recently been experimentally identified for relatively 
small systems of eight atoms~\cite{Greiner}. Previous theoretical studies 
investigated interaction effect in continuous-space bosons within the Gross-Pitaevskii theory~\cite{Lugan, Piraud}.

The theoretical tool we employ to identify localized states is the 
analysis of the energy-level spacing statistics familiar from quantum chaos 
and random matrix theories~\cite{Metha,Haake}. In this framework, one discerns 
delocalized ergodic states from  localized states by identifying the 
Wigner-Dyson statistical distribution and the Poisson distribution of the 
level spacings, respectively. 
This approach has been commonly adopted in studies on single-particle Anderson 
localization in discrete lattice models~\cite{Shklovskii,Hofstetter,Milde,Schweitzer}, 
and more recently also in continuous-space (single-particle) models relevant 
for cold-atoms experiments~\cite{Fratini,Fratini2}. 
Chiefly, this approach has been established as one of the most sound criteria to 
identify many-body localized phases~\cite{Oganesyan}, since it allows one to identify the breakdown of 
ergodicity independently of the specific mechanism causing localization, including, e.g, localization in Fock space.
 In this context, it has 
been applied to one-dimensional discrete systems, including spin 
models~\cite{Pal,Cuevas,Luitz,Laumann}, spinless fermion models~\cite{Oganesyan,Naldesi}, 
and recently also to bosonic models~\cite{Sierant,Sierant2}. Here, we apply it to a 
continuous-space few-boson model.

The first step we take is to determine, in the single particle case, the disorder 
intensity required to observe the statistics of nonergodic systems 
(i.e. the Poisson level-spacing distribution) for a linear system size that is 
feasible for our computational approach -- and, as a matter of fact, also for 
cold-atom experiments -- in a sufficiently broad low energy region of the 
spectrum. Then, we analyze if and how inter-particle interactions affect 
the localization in the two and in the three-particle cases, 
again analyzing in an energy resolved manner the level-spacing statistics. 
This is obtained via large-scale matrix diagonalization calculations of 
the two-body and three-body Hamiltonians represented in the Fock space 
corresponding to a suitably chosen basis. A broad range of interaction 
strengths is considered, ranging from the noninteracting limit, up to strong 
interaction strengths in the broad vicinity of the strongly-interacting limit 
where the bosons fermionize, meaning that the ground-state energy approaches the value corresponding to
 a noninteracting fully-polarized fermion model. 
Furthermore, we characterize how the ground-state energy evolves in the 
crossover between the noninteracting and the strongly-interacting limits.

The main result we report is that, for a reasonably broad low-energy 
portion of the spectrum,  the contact interaction does not induce a 
shift from Poisson to Wigner-Dyson statistics; this statement holds 
for the full interaction range we consider. This indicates that many-body localization
 can occur, at low energy, also in continuous-space models, at least in the few-body 
setup and for the contact interaction model considered here.
The rest of the article is organized as follows:
in Section~\ref{secmodel} we describe the continuous-space model with speckle disorder 
and the computational approach we adopt, analyzing in particular the convergence of 
the energy levels as a function of the basis size.
Section~\ref{secsingleparticle} focuses on a single particle in the speckle disorder, 
analyzing the spatial structure of the eigenstates and the level-spacing statistics.
The results for two-boson and three-boson systems are presented in Section~\ref{secfew}.
Our conclusions and the future perspectives are reported in Section~\ref{secconc}.
\section{The Hamiltonian}
\label{secmodel}
In the general case, the model we consider consists in $N$ identical bosons of 
mass $m$ in a one-dimensional box of size $L$, with a random external field $V(x)$ 
that describes a blue-detuned optical speckle field~\cite{Goodman}. The Hamiltonian reads,
\begin{equation}
\label{Eq.1:ham}
\mathcal{H}=\sum_{i=1}^N \left(-\frac{\hbar^2}{2m}\frac{\partial}{\partial x_i}+V(x_i)\right)
+\sum_{i<j}^N v(|x_i-x_j|).
\end{equation}
The variables $ x_i$, with the particle label $i=1,\,...\,N$, indicate the 
particle coordinates. Hard-wall boundary conditions are considered, meaning 
that the wave functions vanish at the system boundaries. $v(|x_i-x_j|)$ indicates 
a zero-range two-body  interaction potential between particles $i$ and $j$, 
defined as,
\begin{equation}
v(|x_i-x_j|)=g\delta(|x_i-x_j|)\,.
\end{equation}
The coupling parameter $g$, which fixes the interaction strength, is related 
to the one-dimensional scattering length, $a_0$, as $g=-\hbar^2/(m a_0)$. In 
this work, we consider a repulsive interaction, $g\geqslant 0$.
The one-dimensional Hamiltonian~(\ref{Eq.1:ham}) accurately describes ultracold 
gases in one-dimensional waveguides with a tight radial confinement, and the interaction parameter $g$ can be tuned 
either by varying the radial confining strength and/or tuning the three-dimensional 
scattering length via Feshbach resonances~\cite{Olshanii}.

The external field $V(x)$ describes the potential experienced by alkali atoms 
exposed to optical speckle fields. 
Such fields are generated when coherent light passes through a rough (semitransparent) surface. 
An efficient numerical algorithm to create speckle fields in computer simulations 
has been described elsewhere~\cite{Huntley,Modugno,Okamoto,Pilati}, and we refer the readers 
interested in more details about the algorithm to those references.

Fully developed speckle fields in large systems are characterized by an exponential 
probability distribution of the local intensities $V$, which reads $P(V)=\exp(-V/V_0)/V_0$ 
for $V\geqslant 0$, and $P(V)=0$ for $V<0$~\cite{Goodman}. 
Here, $V_0\geqslant 0$ is the average intensity of the field, and coincides with its 
standard deviation. $V_0$ is therefore the unique parameter that characterizes the 
disorder strength.

The two-point spatial correlation function of local intensities of a speckle 
field depends on the distance $d$ between two given points and reads~\cite{Modugno}, 
\begin{equation}
\Gamma(d)={\langle V(x+d)V(x)\rangle\over V_0^2}-1=[\sin(d\pi/\ell)/(d\pi/\ell)]^2 \,.
\end{equation}
Here, the brackets $\langle \cdots\rangle$ indicate spatial averages. Notice that in a 
large enough system, the speckle field is self-averaging, meaning that spatial averages 
can be replaced by averages of local values over many realizations of the speckle field.
The length scale $\ell$ is related to the inverse of the aperture width of the 
optical apparatus employed to create the optical speckle field and to focus it 
onto the atomic cloud.
It characterizes the typical distance over which the local intensities loose 
statistical correlations, or, in other words, the typical size of the speckle grains.
In the following, we will use this spatial correlation length as unit of 
lengths, setting $\ell=1$.
This length scale also allows one to define a characteristic energy scale, often 
referred to as correlation energy, which reads $E_c=\hbar^2/(m\ell^2)$. This quantity 
will be used in the following as the unit for energies, unless explicitly stated. 
The interaction parameter $g$ will be expressed in units of $\hbar^2/(\ell m)$.
%
%
%


\begin{table}[t]
\centering
\begin{tabular}{|c|c|c|c|c|c|}
\cline{1-6}
$E_{\rm max}$   & $M$     & $D_{MB}$     & $E_{GS}(g=0)$ & $E_{GS}(g=1)$ & $E_{GS}^{h.o.}(g=1)$\\ \cline{1-6}
20& 142 &  7941     & 7.0266& 7.7813 & 1.3249 \\ 
40& 201 &  15889    & 6.8818& 7.5954 & 1.3191 \\ 
60& 246 &  23836    & 6.8328& 7.5403 & 1.3167 \\ 
100& 318 &  39747   & 6.8309& 7.5338 & 1.3143 \\ 
120& 348 &  47697   & 6.8308& 7.5319 & 1.3136 \\ \cline{1-6}
\end{tabular}
\caption{Convergence of the ground state energy for $N=2$: for a given 
speckle and in the noninteracting case (forth column) and for $g=1$ (fifth column); 
for an harmonic potential and $g=1$ (sixth column) that should tend to the exact 
value $E_{GS}^{h.o.}\cong1.30675$~\cite{Busch}. $E_{\rm max}$ was used to truncate the many-body 
Hilbert space using an energy criterion which required $M$ single-particle states 
and it corresponds to a many-body Hilbert space dimension $D_{MB}$.
The system size is $L=100\ell/\sqrt{2}$ and in the cases with a harmonic trap we have 
set $\sqrt{\hbar/(m\omega)}=\ell$, i.e., $\hbar \omega=E_c$. For the speckle potential $V_0=50 E_c$.}
\label{table1}
\end{table}

The few-body problem is solved by direct diagonalisation of the 
second-quantized many-body Hamiltonian in a truncated many-body 
basis~\cite{raventostuto}. 
The  Hamiltonian (\ref{Eq.1:ham}) is written in second-quantization as the sum 
of three terms:
\begin{equation}
\hat{H}=\hat{K}+\hat{V}+\hat{v}\,.
\end{equation}
Each term  can be defined using the standard creation and annihilation operators, 
$\hat{a}^\dagger_i$ and $\hat{a}_j$, that fulfill the bosonic commutation relations 
$[\hat{a}_j,\hat{a}^\dagger_i]=\delta_{i,j}$. 
These operators create or annihilate bosons in the single-particle states, which 
are taken as the eigenstates of the free particle moving in a one dimensional box 
with hard walls. The box size is chosen large enough compared to the spatial 
correlation length $\ell$. 
The many-body Fock states are defined in terms of the creation operators as
\beq
\label{Fockbasis}
\ket{n_1,\, ... \, ,n_M}=\frac{(\hat{a}^{\dagger}_1)^{n_1} \dots 
(\hat{a}^{\dagger}_M)^{n_M}}{\sqrt{n_1 !\, ... \, n_M !}} \ket{\rm vac},
\eeq
where $ \ket{\rm vac}$ is the vacuum state and the quantum numbers $n_k$ 
indicate the number of bosons in the single-particle state $k$. In all computations, 
we have $\sum_{k=1}^M n_k=N$. 
In this basis, the kinetic energy, $\hat{K}$, has diagonal form, 
\begin{equation}
\hat{K}=\sum_{k} \frac{k^2 \pi^2}{2L^2}\hat{a}^\dagger_k \hat{a}_k\,.
\end{equation}
The speckle potential reads
\begin{equation}
\hat{V}=\sum_{k,j} V_{k,j}\hat{a}^\dagger_k \hat{a}_j\,,
\end{equation}
with
\begin{equation}
V_{k,j}=\int_{-\frac{L}{2}}^{\frac{L}{2}} dx \,\,\, \psi_k^*(x)V(x)\psi_j(x)\,,
\end{equation}
and, 
\beq
\psi_k(x)=\sqrt{\frac{2}{L}}\sin\left(\frac{k\pi}{L}(x+\frac{L}{2})\right) \,.
\eeq
These integrals are determined via numerical quadrature based on the composite 
five-point Bode's rule, using a sufficiently fine grid so that the residual numerical 
error due to the discretization is negligible.\\
The interaction term reads
\begin{equation}
\hat{v}=\frac{g}{2}\sum_{i,j,k,l} v_{ijkl}\hat{a}^\dagger_i\hat{a}^\dagger_j \hat{a}_k\hat{a}_l,
\end{equation}
with
\begin{equation}
\begin{split}
v_{ijkl}=\frac{1}{2L}\left(-\delta_{i,j+k+l}+\delta_{i,-j+k+l}\right.
\\
+\delta_{i,j-k+l}-\delta_{i,-j-k+l}+\delta_{i,j+k-l}
\\
\left. -\delta_{i,-j+k-l}-\delta_{i,j-k-l}+\delta_{i,-j-k-l}\right).
\end{split}
\end{equation}
%
We follow the criterion to truncate the many-body basis discussed in 
Ref.~\cite{Plodzien}, where its efficiency in terms of computing resources 
has been highlighted. The many-body basis is built including all states 
with a kinetic energy equal or smaller than a given threshold 
$E_{\textrm{max}}$.  The minimal number of 
single-particle modes $M$ required  to include  all such many-body
 states is retained. The energy threshold $E_{\textrm{max}}$ represents an 
algorithmic parameter whose role has to be analyzed. In fact, while the 
computation is exact in the $E_{\textrm{max}}\rightarrow \infty$ limit, a 
residual truncation error might occur for  finite $E_{\textrm{max}}$ value. 
Once the number of bosons and energy threshold are fixed we can build the 
corresponding many-body Fock basis. Then we construct the Hamiltonian in this 
Fock basis and diagonalize its low energy part by means of the ARPACK 
implementation of Lanczos method. 

\begin{figure}[t]
\centering
\includegraphics[width=\columnwidth]{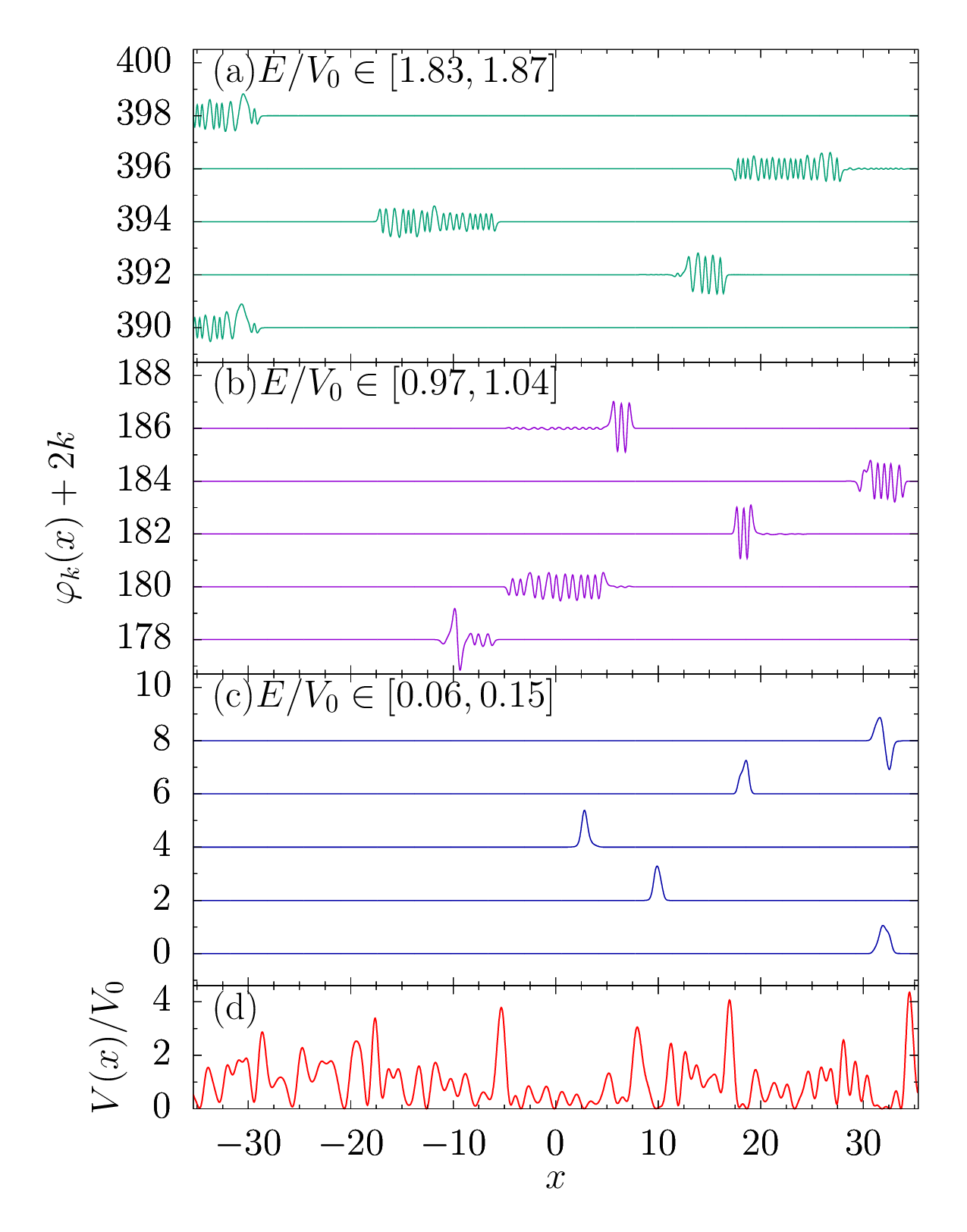}
\caption{In panels (a-c), we present some eigenfunctions of the 
speckle potential, $V(x)$, from bottom to top, in order of increasing 
energy in the energy ranges written in the panels. A realization of the 
speckle potential with $V_0=50 E_c$ is shown in panel (d). The system 
size is $L=100\ell/\sqrt{2}$.}
\label{Fig:1}
\end{figure}

Table~\ref{table1} reports the analysis of the convergence with the energy 
truncation parameter $E_{\textrm{max}}$ for a few representative setups. $D_{MB}$ 
in this table indicates the number of states in the many-body basis set. 
Specifically, we consider the ground-state energy of two bosons in the 
noninteracting case ($g=0$) and with a relatively strong interaction ($g=1$). 
Here the disorder strength is set to $V_0=50 E_c$. One notices that with the largest 
basis set the residual truncation error is much smaller than $0.1\%$. While 
the truncation effect becomes somewhat larger at higher energies, we consider 
in this work an energy range where this effect is negligible. 
An estimate of the accuracy of our numerical procedure can be obtained 
by considering the case of two interacting bosons trapped in a harmonic potential, 
which was exactly solved in Ref.~\cite{Busch}. We choose a harmonic oscillator of 
length $\ell$, which is the typical size of the minima in the speckle potential, 
within our finite box of size $L=100\ell/\sqrt{2}$. For an interaction strength 
$g=1$ we reproduce the exact results up to the second decimal. This provides a 
reasonable estimate of the accuracy of our method.   
Furthermore, we mention here that the results of the analysis of the 
(ensemble averaged) level-spacing statistics are less sensitive to the 
truncation error than the individual energy level of a single realization 
of the speckle field.
%
%
\section{Localization in the single-particle case}
\label{secsingleparticle}
In this Article, the occurrence of the localization, i.e. nonergodic behaviour, is inspected 
by analyzing the statistics of the energy-level spacings.
As a preliminary step, we start by analyzing the general features of 
the single-particle modes.
Fig.~\ref{Fig:1} displays the single-particle wave-functions at low, intermediate, and 
relatively high energies, for a given realization of a speckle field 
of intensity $V_0=50 E_c$. 
The low-energy states are located in the Lifshitz tail of the density 
of states. Here, localization typically occurs in rare regions where 
the disorder creates a deep well confined by tall barriers.
In fact, we observe that the spatial extent where these low-energy 
states have large amplitude is typically of the order of the disorder 
correlation length $\ell$, meaning that they are indeed localized 
in a single well of the speckle field.
However, this spatial extent rapidly increases as a function of the energy, becoming 
significantly larger than $\ell$.
On a qualitative level, this effect can be observed in Fig.~\ref{Fig:1}, noticing that the 
states at intermediate and at relatively high energies have large amplitude in several 
wells of the speckle potential.
To quantify this spatial extent, we compute the participation ratio, which is 
defined as $P_k = 1/\int \mathrm{d} x \left|\varphi_k(x)\right|^4$.
For the low-energy states in the Lifshitz tail, we find, again for 
$V_0=50 E_c$, $\langle P_k \rangle\simeq \ell$, indeed corresponding to trapping 
in a single deep well. Here the brackets $\langle \cdot \rangle$ indicate the 
average over many realizations of the speckle field.
Instead, for states with energies above the average speckle-field intensity, 
e.g., with energy $E \simeq 2V_0$, the spatial extent is 
$\langle P_k \rangle \simeq 5\ell$, and it reaches 
$\langle P_k \rangle \simeq 11\ell$ at $E \simeq 3V_0$.
At even higher energies the participation ratio is of the order of the system 
size (here $L=100\ell/\sqrt{2}$) and finite-size effects due to the box 
become dominant. 
At these energies the single-particle states are weakly affected by the disorder, 
since in the finite system the speckle field typically develops only moderately 
high peaks, as opposed to an infinite system where a sufficiently high peak would 
always occur given that the speckle potential has no upper bound.
Clearly, these finite-size effects have to be avoided (see also the discussion 
on the analysis of the level-spacing statistics reported below).
The choice of inspecting that localization occurs in a sufficiently small length 
scale and in a reasonably broad portion of the energy spectrum, here taken of 
the order of the average speckle field intensity $V_0$, is motivated by the 
aim to address, in the second step, the effects of interparticle interactions. 
These will indeed induce population of relatively high energy states even when 
noninteracting bosons would occupy only deeply localized low-lying modes. 
In fact, previous lattice calculations predicted that in one dimension the 
localization length of two interacting particles can be significantly larger 
than the spatial extent of the single particle states~\cite{shepelyansky}. 
In three-dimensional (lattice) systems two-particle interactions could even 
induce complete delocalization~\cite{orso}.
One should also consider that in cold-atom experiments the atomic energy distribution 
is inevitably broadened by thermal excitations, by interactions, and by the finite 
spatial spread of the atomic cloud, meaning that localization effects cannot 
be observed if only very few low-energy states are spatially localized.
In this regard, it is worth mentioning that if one aims at experimentally 
visualizing the exponentially decaying tails of the single-particle states, 
a feature that characterizes Anderson localized systems, it is convenient to 
consider rather weak disorder $V_0 \approx E_c$, since in this regime the 
spatial extent is much larger than the typical well size.
For example, for a speckle-field intensity $V_0 = E_c$ we find 
$P_k \simeq 20 \ell$ at $E=3V_0$. In this case, in order to avoid 
finite-size effects in the participation-ratio calculation (and 
also in the analysis of the level spacing statistics discussed below), 
a system size larger than $L = 1000\ell/\sqrt{2}$ is required.
Such system sizes cannot be addressed with the computational technique we 
employ for interacting systems, therefore in the following we consider larger 
disorder strengths where finite size effects can be more easily suppressed.
Anderson localization in strong speckle disorder has been investigated also in Ref.~\cite{Hilke}.

It is also worth emphasizing that in an infinite one-dimensional system where the disorder
has no upper bound (like the blue-detuned speckle potential), a classical particle is localized 
at any energy $E$, just like a quantum particle in the same setup~\cite{Gang4}. Indeed, 
a position in space where $V(x)>E$ always occurs, prohibiting the particle 
from exploring the whole configuration space, resulting in nonergodic behavior.
This scenario is different from the one that occurs in two-dimensional~\cite{Weinrib} 
and in three-dimensional systems, where classical particles in a speckle potential 
are trapped only if their energy is lower than a finite threshold; above this energy 
threshold a (classical) percolation transition takes place. In particular, in three 
dimensional speckle potentials the classical percolation threshold turns out to be a 
tiny fraction of the average speckle-potential intensity $V_0$~\cite{Pilati,Jendrzejewski}. 
The mobility edge, i.e. the energy threshold that in three dimensional quantum 
systems separates localized states from extended states, is typically much larger 
than this classical percolation threshold, meaning that in a broad energy range 
particles are trapped purely by quantum mechanical effects.
In the one-dimensional setup considered here, instead, both quantum and classical 
particles are localized at any energy in the infinite-size limit, meaning that 
classical and quantum trapping mechanisms cannot be rigorously separated.

\begin{figure}[t]
\centering
\includegraphics[width=\columnwidth]{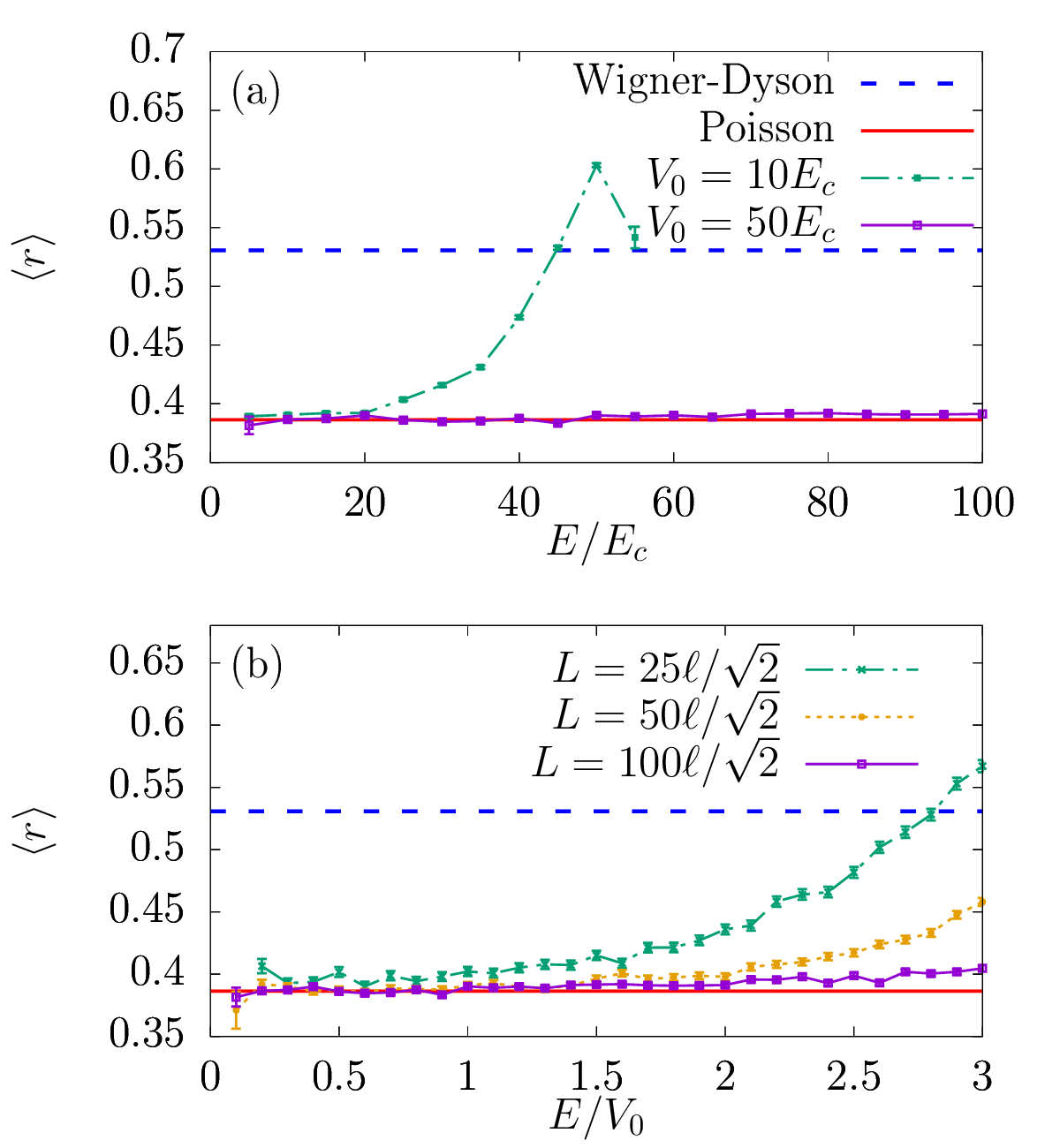}
\caption{The value of $\langle r \rangle$ averaged over $2000$ different 
speckle instances is shown. In (a) we compare two cases with different speckle 
intensity, $V_0$, and in (b) with different system sizes. In panel (a) 
$L=100\ell/\sqrt{2}$ and in panel (b) $V_0=50 E_c$. The distribution in energy 
is computed using energy windows $\Delta E=5 E_c$. The diagonalization was 
performed using a Hilbert space dimensions of 1000.
}
\label{Fig:2}
\end{figure}

The analysis of the statistical distribution of the spacings between consecutive energy 
levels allows one to discern localized (i.e., nonergodic) states from delocalized ergodic states. 
Specifically, localized states are associated to the Poisson distribution of the level 
spacings, while delocalized states are associated to the Wigner-Dyson distribution typical of 
random matrices. An efficient procedure to identify these two distributions consists in 
determining the average over a large ensemble of speckle fields of the following
ratio of consecutive level spacings~\cite{Oganesyan}:
\begin{equation}
\label{eq:ratio}
r_i=\min\left\{\frac{E_{i+1}-E_{i}}{E_{i}-E_{i-1}},\frac{E_{i}-E_{i-1}}{E_{i+1}-E_{i}}\right\}.
\end{equation}
Notice that the ensemble averaging we perform, indicated as 
$\langle r \rangle$, is energy resolved, meaning that only states within a narrow 
energy window are considered. This allows us to address possible scenarios where 
both localized states and delocalized states occur, but in different sectors of 
the energy spectrum.
The Poisson distribution translates to the ensemble average  $\langle r \rangle\cong 0.38629$, 
while the Wigner-Dyson distribution translates to $\langle r \rangle\cong 0.53070$~\cite{Atas}.

As discussed above, the scaling theory of Anderson localization~\cite{Gang4} predicts that 
in infinite one-dimensional disordered systems
the localization occurs for any amount of disorder, even if this amount 
is vanishingly small. However, in finite-size systems the localization length 
might be comparable to the system size, hindering the observation of the 
Poisson distribution corresponding to localized systems.
This effect is particularly relevant if the disorder is weak or if the energy window 
under consideration is high, since the localization length is large in these regimes,
 as previously discussed.
It is, therefore, pivotal for our purposes to identify a disorder strength and 
an energy range where the Poisson statistics  can be observed in a system size that 
is feasible for our computational approach for interacting systems.
Fig.~\ref{Fig:2} displays the energy-resolved analysis of the level-spacings 
statistics for a few representative setups of the optical speckle field.
Specifically, panel (a) shows $\langle r \rangle$ versus $E/E_c$ for a fixed 
system size and different disorder strengths, while panel (b) shows data 
corresponding to different system sizes at a fixed disorder strength.
One observes that, at low energy, the $\langle r \rangle$ values precisely agree 
with the prediction for the Poisson distribution, indicating that the low-energy 
states are localized on a sufficiently small length scale. However, 
significant deviations occur at higher energies. We attribute them to the 
finite-size effect discussed above. In fact, one observes that for larger 
system sizes the Poisson-distribution result extends to higher energies. This 
finding is consistent with the expectation that in an infinite system the 
whole energy spectrum would be localized.
In the following, we will consider the system size $L=100\ell/\sqrt{2}$ and the 
disorder strength $V_0 =50E_c$, where the $\langle r \rangle$ values precisely  
correspond to the statistics of localized systems in a reasonably broad 
energy range $0< E\lesssim 100E_c$. Notice that the upper limit is twice as large 
as the average speckle-field intensity $V_0$.
\begin{figure}[t!]
\centering
\includegraphics[width=0.9\columnwidth]{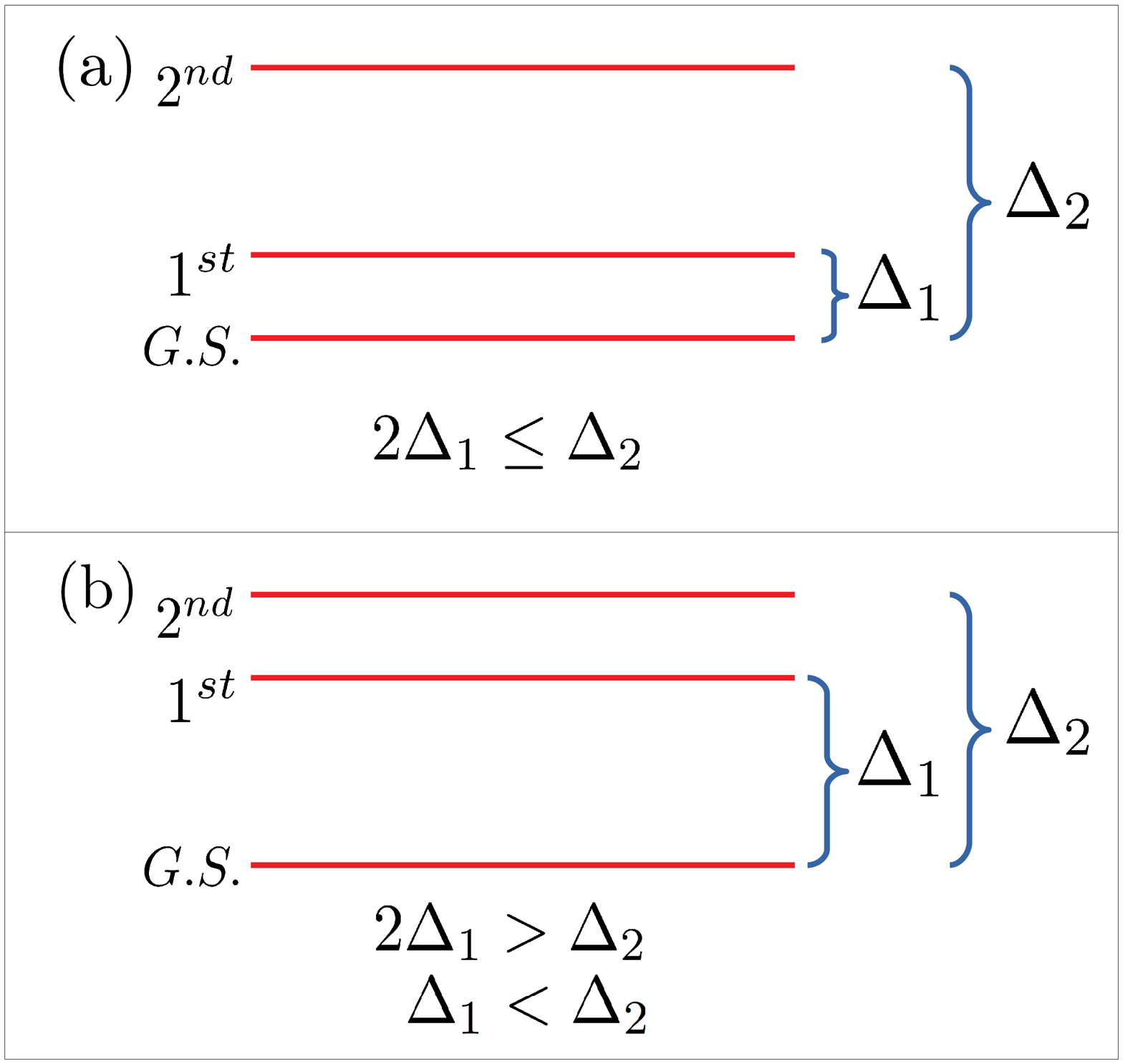}
\caption{The lowest single-particle eigenenergies for a given speckle potential determine 
the lowest energy levels for the noninteracting many-body system. There are two possible 
situations: (a) a small gap between the single-particle ground state ($G.S.$) and the 
first excited state ($1^{\rm st}$); and (b) a small gap between the first excited state 
and the second excited state ($2^{\rm nd}$).}
\label{Fig:8}
\end{figure}

It is worth pointing out that the linear system size of typical cold-atom experiments 
performed with optical speckle field is comparable to the system size considered 
here; it ranges from a few tens to around thousand times the speckle correlation 
length $\ell$. Therefore, this analysis also serves as a guide for experiments on localization phenomena 
in atomic gases.

While the next section is devoted to systems with $N=2$ or $N=3$ interacting bosons, we 
address here the special case of $N>1$ noninteracting particles. Clearly, the system 
properties in this case can be traced back to the single-particle problem. However, as 
we discuss here, special care has to be taken in order to correctly extract the 
correct level spacing statistics.

\begin{figure}[t]
\centering
\includegraphics[width=\columnwidth]{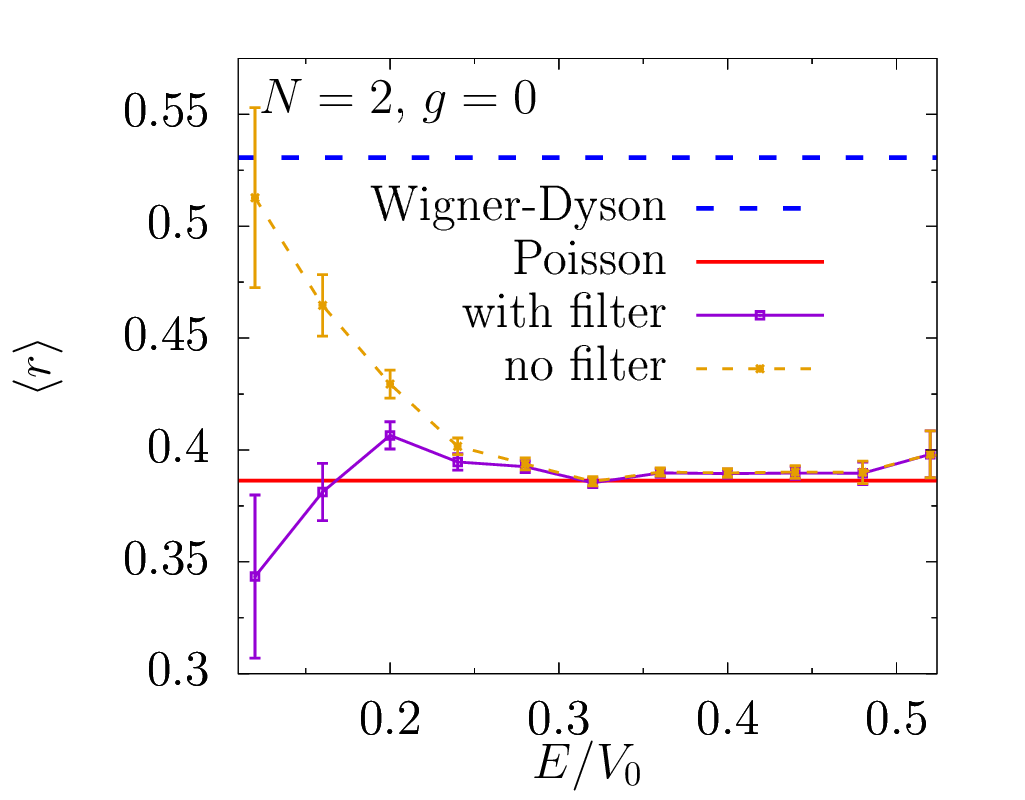}
\caption{Mean value of $r$ as a function of $E/V_0$ distributed in energy computed with and 
without a  randomness filter  for the noninteracting two-boson system. 
The filter removes the values $r_i \geqslant 0.999$.}
\label{Fig:9}
\end{figure}

In fact, in certain circumstances, the $N$-boson energy-level spacings in the 
noninteracting limit  take specific, nonrandom values.
%
%
%
For a given realization of the speckle potential, we can distinguish two possible 
scenarios, depicted in the two panels of Fig.~\ref{Fig:8}, depending on the relative 
distances of the first and of the second single-particle levels from the single-particle ground-state; they are 
indicated below as $\Delta_1$ and $\Delta_2$, respectively. For the 
scenario displayed in panel (a) of Fig.~\ref{Fig:8}, where $2\Delta_1<\Delta_2$, 
the three lowest-energy eigenstates of the noninteracting N-boson system are
\begin{equation}
\begin{gathered}
\ket{E_0}=\ket{N,0,...,0},
\\
\ket{E_1}=\ket{N-1,1,0,...,0},
\\
\ket{E_2}=\ket{N-2,2,0,...,0},
\end{gathered}
\end{equation}
and their associated energies are (see Fig.~\ref{Fig:8})
\begin{equation}
\begin{gathered}
E_0=N E_{GS},
\\
E_1=N E_{GS}+\Delta_1,
\\
E_2=N E_{GS}+2\Delta_1,
\end{gathered}
\end{equation}
where $E_{GS}$ is the single-particle ground state energy.
In this situation, the value of $r_1$ associated to the lowest energy of the system is
\begin{equation}
r_1=\frac{E_1-E_0}{E_2-E_1}=\frac{\Delta_1}{\Delta_1}=1.
\end{equation}
One notices that this does not randomly fluctuate for different speckle-field realizations.\\
In the second scenario (see panel (b) of Fig.~\ref{Fig:8}), where $2\Delta_1>\Delta_2$, 
the three lowest-energy eigenstates of the system are
\begin{equation}
\begin{gathered}
\ket{E_0}=\ket{N,0,...,0},
\\
\ket{E_1}=\ket{N-1,1,0,...,0},
\\
\ket{E_2}=\ket{N-1,0,1,0...,0},
\end{gathered}
\end{equation}
and their associated energies are
\begin{equation}
\begin{gathered}
E_0=N E_{GS},
\\
E_1=N E_{GS}+\Delta_1,
\\
E_2=N E_{GS}+\Delta_2.
\end{gathered}
\end{equation}
Therefore, we have
\begin{equation}
r_1=\frac{E_2-E_1}{E_1-E_0}=\frac{\Delta_2-\Delta_1}{\Delta_1}.
\end{equation}
This is a random variable which depends on the level spacings, and one expects 
it to follow the Poisson (or eventually the Wigner-Dyson) distribution.

If the data emerging from both scenarios are included in the ensemble average, 
one obtains, in the low-energy regime,  $\langle r \rangle$ values with an upward bias, 
therefore deviating from the Poisson statistics even in setups where the 
single-particle modes are localized on a short lenth-scale. This effect, displayed in Fig.~\ref{Fig:9}, 
for the representative setup with $N=2$, $L=100\ell/\sqrt{2}$ and $V_0=50 E_c$, should 
not be associated to a delocalization phenomenon. 
For this reason, in our calculations with $N>1$ noninteracting particles we introduce 
a filter that removes the $r_i$ values which are numerically indistinguishable from 
$r_i=1$, i.e. the data corresponding the the first scenario described above.
With this filter, the ensemble-averaged $\langle r \rangle$ values agree with the 
Poisson distribution result within statistical uncertainties (see Fig.~\ref{Fig:9}). 
As expected, the filter has no effect at moderate to high energies.
It is worth emphasizing that this effect occurs only for noninteracting particles. As soon 
as $g>0$, the many-body state is a superposition of many basis states; so, the two scenarios 
described above do not apply, and  the $r_i$ values randomly fluctuate for different speckle 
field realizations.
\section{Two and three interacting bosons}
\label{secfew}
\begin{figure}[t]
\centering
\includegraphics[width=\columnwidth]{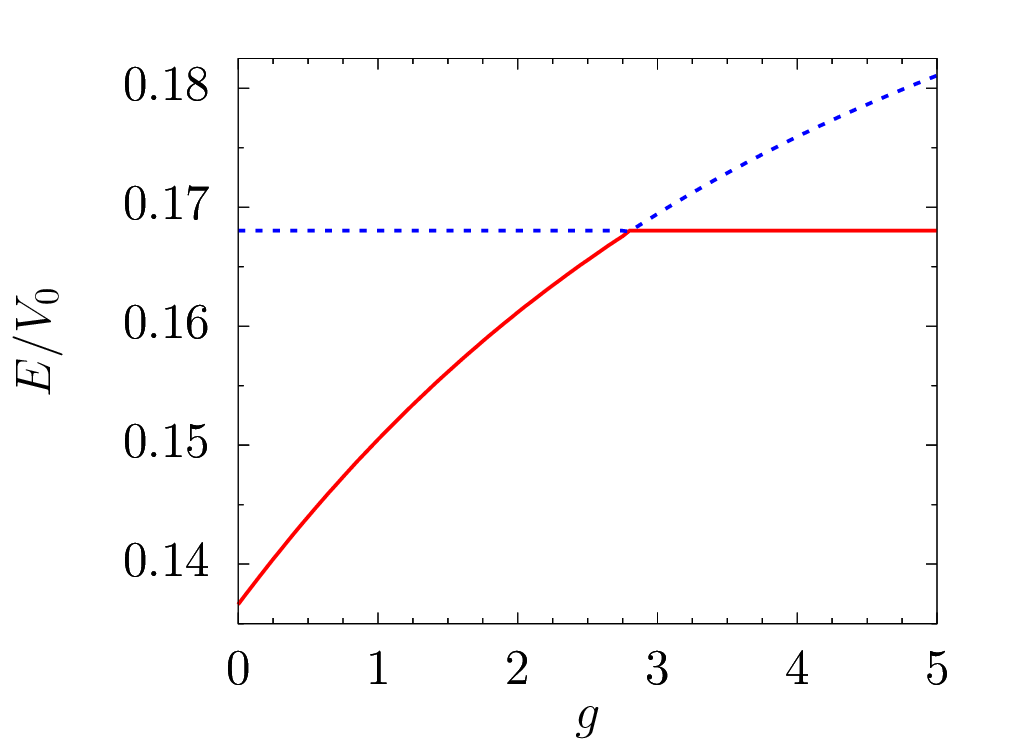}
\caption{Energies of the ground state (red solid line) and first excited state (blue short-dashed line) 
of the system of $N=2$ bosons in the speckle potential of Fig.~\ref{Fig:1}, panel (b), 
as a function of the interaction strength $g$. This figure is obtained with $M=636$, which 
results in a $D_{MB}=159069$ for an $E_{\rm max}=400$.}
\label{Fig:3}
\end{figure}

We start the discussion on the interacting few-boson setup with a qualitative analysis 
of the  interaction effect on the ground-state energy.
Specifically, we consider $N=2$ bosons in a speckle field of intensity $V_0=50 E_c$, 
in a $L=100\ell/\sqrt{2}$ box. As discussed in the previous section, in this setup 
the single-particle modes are spatially localized in a broad energy-range $0<E\lesssim 2V_0$.\\
In the noninteracting limit, the ground state is the Fock-basis state $\ket{2,0,\,...\,,0}$, 
and the corresponding energy equals two times the single-particle 
ground-state energy. In the first excited state, one boson is promoted to the first single-particle 
excited state, obtaining the Fock-basis state $\ket{1,1,\,...\,,0}$. 
The energy levels corresponding to the ground state and to the first excited 
state of a speckle field instance are displayed in Fig.~\ref{Fig:3}, as a function 
of the interaction parameter $g$. One notices that, while the ground-state energy increases 
with $g$, the first excited-state energy is essentially unaltered. This is due to the 
fact that in the excited state the two bosons are localized in far apart wells; 
therefore, the zero-range interaction has an almost negligible effect.
In the strongly interacting limit, $g\rightarrow \infty$, the lowest-energy state 
is $\ket{1,1,\,...\,,0}$. This scenario is similar to the Tonks-Girardeau gas, where bosons 
with infinitely-strong zero-range repulsive interaction can be mapped to a system 
of noninteracting indistinguishable fermions, which occupy different single-particle 
modes due to the Pauli exclusion principle.
Remarkably, the transition between the noninteracting and the strongly-interacting regimes 
is extremely sharp. This effect is due to the long separation between the two lowest-energy 
minima for this realization of the speckle potential. For the speckle field instance 
analyzed in Fig.~\ref{Fig:3}, this sharp crossover 
occurs at $g\approx2.8$. 
Beyond this pseudo-critical 
point the two-boson system is effectively fermionized, meaning that their ground-state energy 
essentially coincides with the one of two identical fermions in the same setup. 
Remarkably, this fermionization occurs at strong but finite values 
of the interaction parameter $g$, as opposed 
to homogeneous systems where bosons fermionize only in the $g\rightarrow \infty$ limit,
which correponds to the standard Tonks-Girardeau gas.
Note that in the speckle instances in which the ground and first excited single 
particle states are localized in the same minima, the fermionization would be smoother, 
in line with, e.g. fermionization in a harmonic potential~\cite{deuretzbacher}.

\begin{figure}[t]
\centering
\includegraphics[width=\columnwidth]{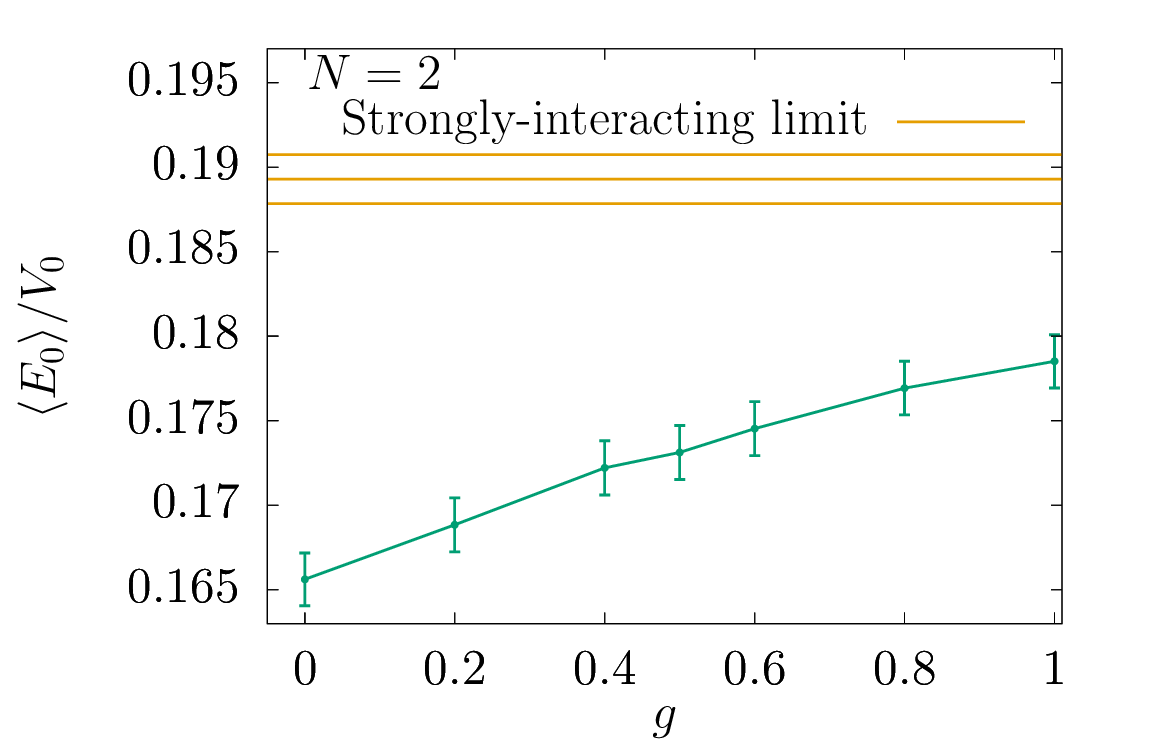}
\caption{Mean value of the ground state energy averaged over $N_s=985$ different 
speckle potentials as a function of the interaction strength $g$ for the two-boson 
system. The error bars are computed as the standard deviation, 
$\sigma_{E_0}=\sqrt{\frac{\langle E_0^2\rangle - \langle E_0\rangle^2}{N_s}}$. 
The speckle realisations used to produce this plot are the same as those used 
in Fig.~\ref{Fig:4}.}
\label{Fig:5}
\end{figure}

\begin{figure}[t]
\centering
\includegraphics[width=\columnwidth]{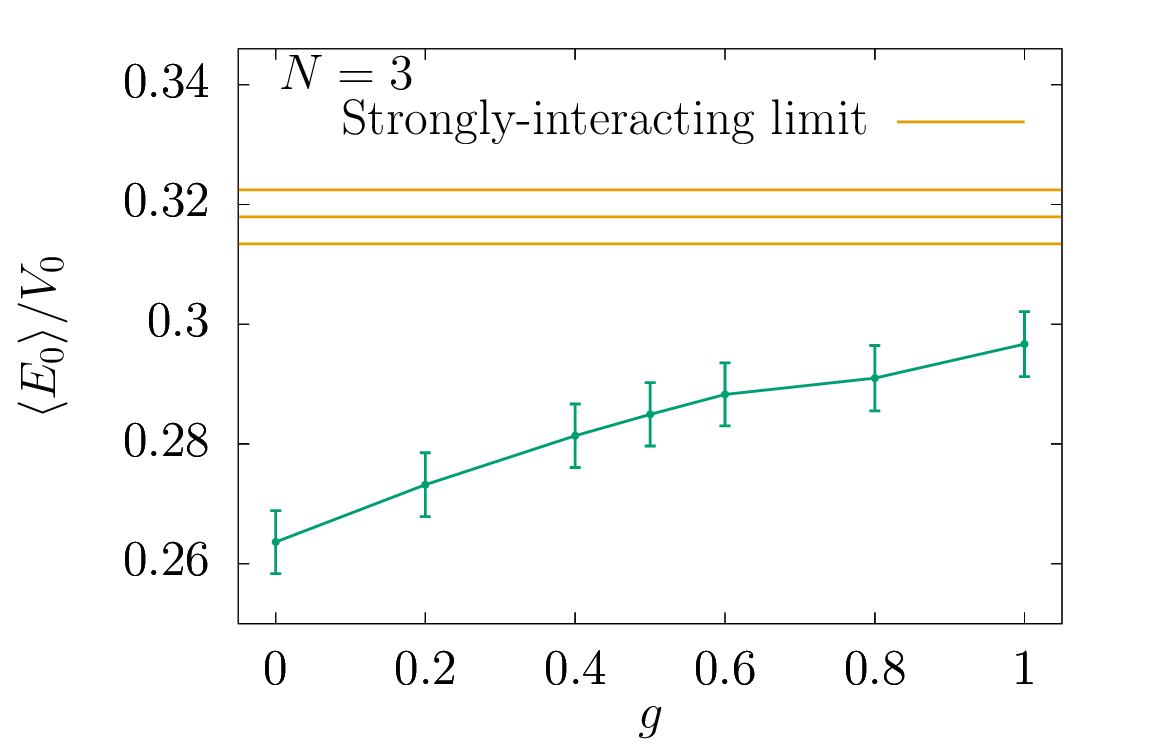}
\caption{Mean value of the ground state energy averaged over $N_s=250$ different speckle 
potentials depending on the interaction strength $g$ for the three-boson system. 
The error bars are computed as the standard deviation, 
$\sigma_{E_0}=\sqrt{\frac{\langle E_0^2\rangle - \langle E_0\rangle^2}{N_{s}}}$. 
The speckle realisations used to produce this plot are the same as those used 
in Fig.~\ref{Fig:6}.}
\label{Fig:7}
\end{figure}

For different speckle field instances, this fermionization transition occurs at different values 
of the coupling parameter $g$. Also the energy levels in the noninteracting limit and in the 
strongly-interacting limit, as well as in the crossover region, randomly fluctuate.
In Fig.~\ref{Fig:5}, the average over many realizations of the speckle field of the 
two-boson ground-state energy is plotted as a function of the interaction parameter $g$. 
Here, we consider interaction strengths ranging from the noninteracting limit to the moderately 
large interaction parameter $g=1$. 
One notices that this interaction strength is sufficient 
to shift the ground-state energy away from the noninteracting-limit result, reaching 
values in fact closer to the strongly-interacting limit --- where the energy of a noninteracting 
identical fermions is reached --- than to the noninteracting limit. The 
same scenario occurs for the $N=3$ boson system, which is analyzed in Fig~\ref{Fig:7}.
In the following, we focus on the interaction regime  $0\leqslant g\leqslant 1$, where any 
interesting interaction effect would take place. Stronger interactions  require extremely 
large basis-set sizes, so that it is not computationally feasible for us to perform 
averages of many realization of the speckle field.
This regime of intermediate interaction strength $g\approx 1$ is, in fact, the one 
where one expects to have more pronounced delocalization effects. Indeed, in the 
strongly-interacting limit the system properties are again determined 
by the single-particle modes. Since the latter are localized for the disorder 
strength considered here, one expects the many-body system to be localized, too.
This type of re-entrant behavior has been observed in the cold-atom experiments on 
many-body localization~\cite{Schreiber}. The experimentalists  indeed found that in the 
strongly-interacting limit the system is many-body localized if the corresponding noninteracting 
system is localized. The experiment was performed with fermions with two spin 
states. In this case, in the strongly-interacting limit the system properties can be mapped to those of 
a fully polarized (noninteracting) Fermi gas, in analogy with the Tonks-Girardeau physics in Bose gases.


\begin{figure}[t!]
\centering
\includegraphics[width=\columnwidth]{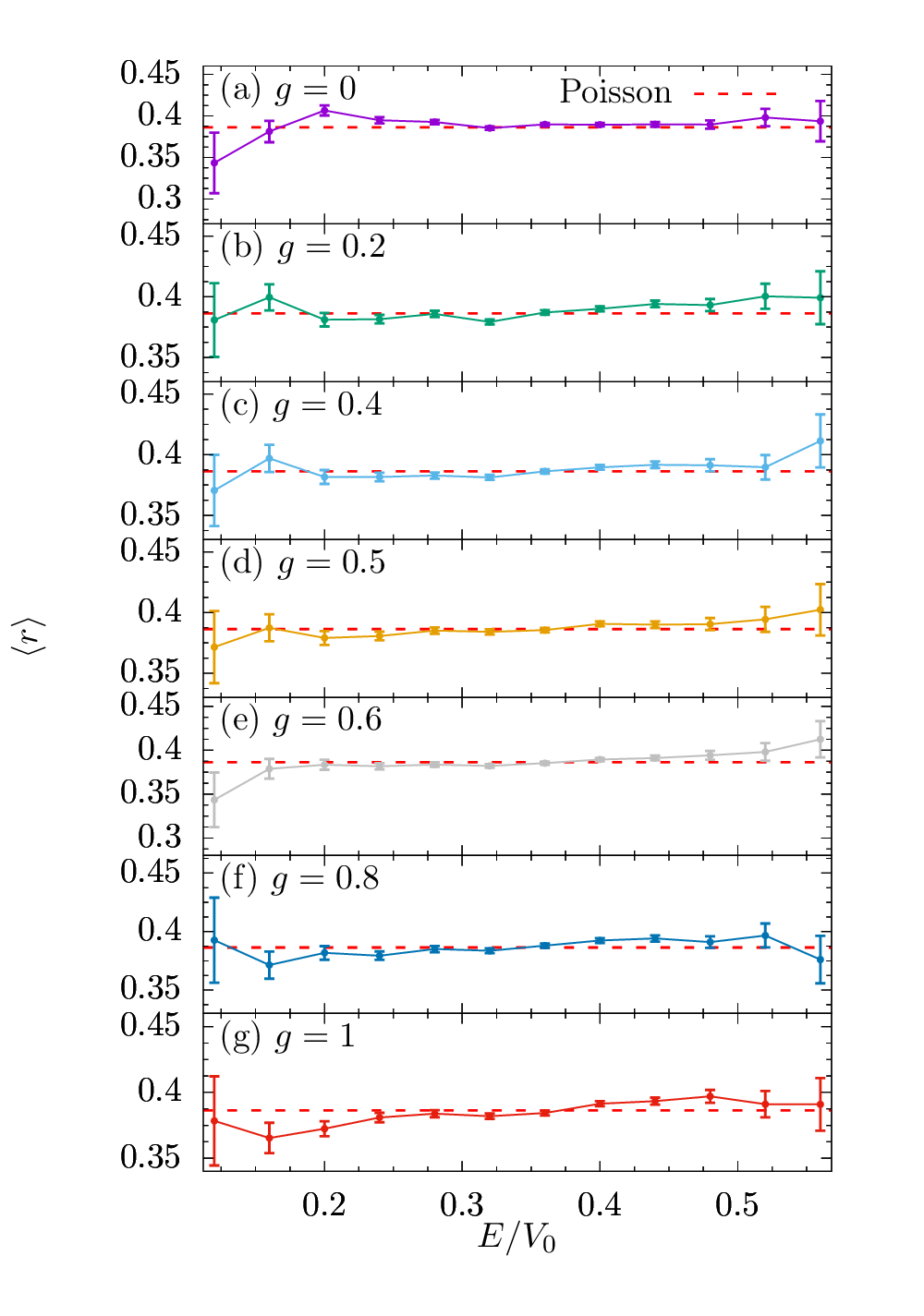}
\caption{Distribution in energy of $\langle r \rangle$ for $N=2$ 
bosons in a 1D box with a speckle potential. The numerical results with different 
interaction strengths $g$, of  a contact potential, are compared with the theoretical 
value that correspond to a Poisson distribution of the energy gaps.} 
\label{Fig:4}
\end{figure}

The analysis of the level-spacings statistics for the interacting two-boson system is 
displayed in Fig.~\ref{Fig:4}. Specifically, we plot the disorder-averaged  $\langle r \rangle$ 
values as a function of $E/V_0$, for different values of the interaction parameter $g$.
The disorder strength $V_0$ and the linear system size $L$ are the ones discussed above and in the previous section.
We focus on the low-energy regime $E\lesssim V_0$. Accurately computing more energy levels 
for many speckle-field instances, in particular, at high energies where larger basis sets 
are required, exceeds our computational resources.

For the computations of Fig.~\ref{Fig:4} the basis sets includes $23836$  states, namely 
the ones with a kinetic energy less or equal to $E_{\mathrm{max}}=60E_c$; this corresponds to 
employing $M=246$ single-particle modes. The disorder ensemble includes $985$ realizations of the speckle field. \\
It is clear that  the $\langle r \rangle$ values are always consistent with the prediction 
corresponding to the Poisson distribution of the level spacings, which is associated to nonergodic systems. 
The statistical uncertainty is larger in the $E\rightarrow 0$ limit due to the low density 
of states in the low energy regime, which reduces the available statistics.
The agreement with the Poisson distribution implies that, for the range of coupling 
constant considered here, the zero-range interaction does not induce delocalization of the two-boson system.
It is possible that a two-body mobility edge, separating low-energy localized states 
from  high energies extended states, would occur at higher energies. However, addressing 
higher energies requires larger computational resources and it is beyond the scope of the present article.

\begin{figure}[t!]
\centering
\includegraphics[width=\columnwidth]{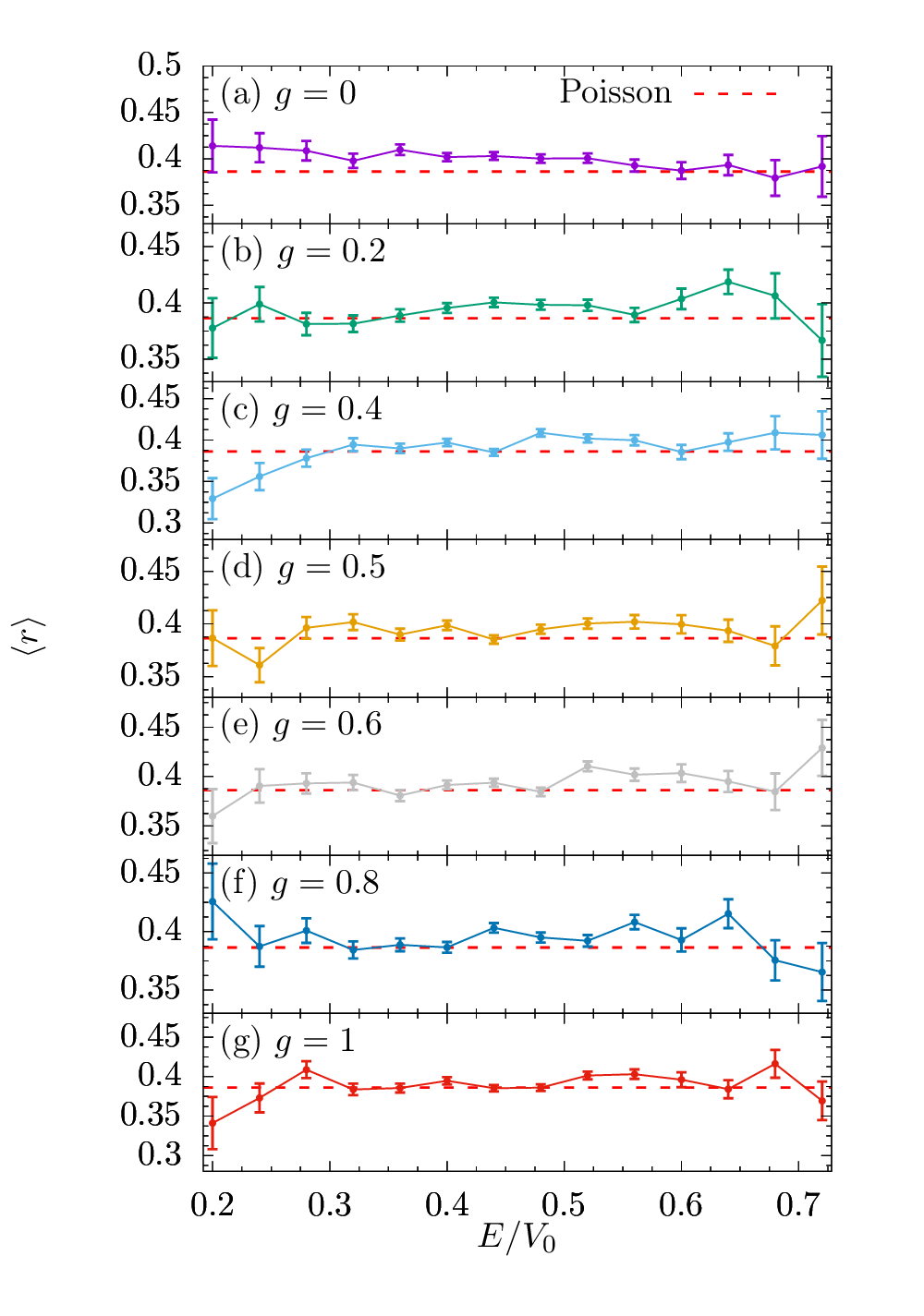}
\caption{Distribution in energy of $\langle r \rangle$ for $N=3$ 
bosons in a 1D box with a speckle potential for different interaction strengths. The 
numerical results are compared with the theoretical predictions corresponding to a 
Poisson distribution of the energy gaps.}
\label{Fig:6}
\end{figure}

The results for the $N=3$ bosons systems are shown in Fig.~\ref{Fig:6}. Here, the basis-set size 
is $117977$, corresponding to the Fock basis states with a kinetic energy less or equal to 
$E_{\mathrm{max}}=12E_c$, in turn implying the use of $M=110$ single particle modes. The 
disorder-ensemble includes $250$ realizations of the speckle field.
We observe that also in the three-boson system localization is, in the low energy regime 
and for the coupling parameters considered here, stable against the effect of zero-range interactions.
\section{Summary and Conclusions}
\label{secconc}
We have performed a computational investigation to establish if and how zero-range repulsive 
interactions can induce ergodic behavior in an otherwise localized system. While previous computational investigations 
on this critical issue addressed discrete-lattice models, we focused here on a model defined in 
continuous space. 
Specifically, we considered a one-dimensional model which describes ultracold 
atoms exposed to random optical speckle patterns, taking into account the structure of the spatial 
correlations of the disorder field. This is the setup that has been implemented in early cold-atom 
experiments on the Anderson localization. 
\\
The computational procedure we employed is based on exact-diagonalization calculations, combined 
with the statistical analysis of the levels-spacings statistics familiar from random matrix theory. 
This is, in fact, one of the most sound criteria commonly employed to identify localized (i.e., nonergodic) 
phases in noninteracting as well as in interacting disordered systems, where it allows to identify 
many-body localized phases.\\
As a preliminary step, we identified the speckle-field intensity required to observe the 
(Poisson) statistics of localized systems in a finite linear system size that is feasible 
for our computational approach and for cold-atom experiments.
Our main finding is that, if two or three interacting bosons move in such a speckle field, 
the localization is stable against zero-range interparticle interactions in a broad 
range of interactions strengths, ranging from the noninteracting limit, up to moderately 
strong interactions half way to the strongly-interacting limit. Addressing even stronger interactions 
is beyond the scope of this article since, on the one hand, it would require larger computational 
resources and, on the other hand, delocalization effects due to interactions are not expected 
in this regime since in the strongly-interacting limit the system properties are determined by the single-particle modes.
Our results are limited to a low-energy regime, of the order of the speckle-field intensity 
$E \lesssim V_0$, where the accuracy of the diagonalization results is under control. It 
is possible that at higher energies two-body or three-body mobility edges would occur. 
We leave this question to future investigations.\\

Previous studies on the possible occurrence of many-body localization in continuous-space 
systems have provided contradictory results. The findings reported here establish that in a 
few-body system  localization can be stable against zero-range interactions in a 
continuous-space models relevant for cold-atom experiments.\\

We conclude by formulating two additional interesting questions, that naturally emerge from 
the previous discussion:  i) would finite-range or long-range interactions  induce delocalization 
in an otherwise localized system?  ii) Would a localization-delocalization transition 
take place if more particles were included? We leave these two question to future investigations.

\vspace*{0.5cm}
\begin{acknowledgments}
We acknowledge financial support from the Spanish Ministerio de Economia y Competitividad 
Grant No FIS2017-87534-P. P.M. is supported by a FI grant from Generalitat de Catalunya.
S.P. acknowledges financial support from the FAR2018 project of the University of Camerino, 
and the CINECA award under the ISCRA initiative, for the availability of high performance 
computing resources and support. S. P. also acknowledges travel support from ICCUB. 

\end{acknowledgments}


\end{document}